# Fast fluorescence lifetime imaging analysis via extreme learning machine

Zhenya Zang,[1] Dong Xiao,[1] Quan Wang[1], Zinuo Li[2], Wujun Xie,[1] Yu Chen,[2] David Day-Uei Li[1,*]

[1]*Department of Biomedical Engineering, University of Strathclyde, Glasgow, G4 0RE, United Kingdom*
[2]*Department of Physics, University of Strathclyde, Glasgow, G4 0NG, United Kingdom*
*\*Corresponding author: david.li@strath.ac.uk*

**Abstract:** We present a fast and accurate analytical method for fluorescence lifetime imaging microscopy (FLIM) using the extreme learning machine (ELM). We used extensive metrics to evaluate ELM and existing algorithms. First, we compared these algorithms using synthetic datasets. Results indicate that ELM can obtain higher fidelity, even in low-photon conditions. Afterwards, we used ELM to retrieve lifetime components from human prostate cancer cells loaded with gold nanosensors, showing that ELM also outperforms the iterative fitting and non-fitting algorithms. By comparing ELM with a computational efficient neural network, ELM achieves comparable accuracy with less training and inference time. As there is no back-propagation process for ELM during the training phase, the training speed is much higher than existing neural network approaches. The proposed strategy is promising for edge computing with online training.

## 1. Introduction

Fluorescence lifetime imaging microscopy (FLIM) has attracted growing interest in biomedical applications such as surgical procedures [1], tumor detection [2, 3], cancer diagnosis [4], and the study of protein interaction networks using Förster resonance energy transfer (FRET) techniques [5]. It can quantitatively investigate local microenvironments of fluorophores by measuring fluorophores' lifetimes. For example, FLIM can observe dynamic metabolic changes in living cells by measuring autofluorescence lifetimes of NAD and NADP. This is utilized to mediate cell fate for diabetes and neurodegeneration research [6]. Fluorescence lifetime is the average time a fluorophore stays excited before releasing fluorescence. The process can be analyzed in the time- or frequency domain. Time-correlated single-photon counting (TCSPC) techniques [7] are more widely used [8-10] due to their superior signal-to-noise ratio (SNR) and precise temporal resolution (in picoseconds) compared with frequency-domain approaches. During data acquisitions, emitted photons are detected by a single-photon detector, wherein a high-precision stopwatch circuit records timestamps of detected photons. The stopwatch circuit generates an exponential histogram, from which the fluorescence lifetime is extracted.

Estimating lifetime parameters is an ill-posed problem with high computational complexity. Numerous algorithms have been developed to quantify lifetimes and relevant parameters. Iterative fitting and optimization approaches were reported to deduce fluorescence lifetimes. A convex optimization method [11] was utilized for high-resolution FLIM, where the accuracy is related to fine-tuned hyperparameters in the cost function. An F-value-based optimization algorithm [12] was used to minimize signal distortion introduced by pile-up effects and the dead-time of single-photon detectors. A Laguerre expansion method [13-15] was reported to speed up least-squares deconvolutions.

On the other hand, non-iterative fitting methods were introduced to reduce computing complexity whilst maintaining high accuracy. A new nonparametric empirical Bayesian framework [16] was adopted for lifetime analysis based on a statistical model, where the expectation-maximization algorithm was employed to solve the optimization problem. A hardware-friendly fitting-free center of mass (CMM) [17-19] algorithm was proposed to deliver fast analysis and has been applied to a flow-cytometry system [20, 21]. Integral equation methods (IEM) [22] were also implemented in FPGA devices to provide real-time analysis. Direction-of-arrivals estimation [23] was adopted to deliver a non-iterative and model-free lifetime reconstruction strategy requiring a few time bins. A histogram cluster method [24] divides histograms into clusters instead of processing histograms pixel-by-pixel, enhancing the analysis speed. However, challenges remain. Firstly, most of these algorithms need a long acquisition time to guarantee the reconstruction fidelity, likely causing photobleaching. A fast algorithm suitable for low photon counts conditions is therefore desirable. Secondly, iterative or probabilistic methods are not portable to hardware, impeding the on-chip computing of TCSPC systems.

Artificial neural networks (ANNs) have been proved promising for FLIM analysis. FLI-NET [25] used a 3-D convolutional neural network (CNN) to analyze bi-exponential decays via a branched architecture. Its compressed-sensing [26] version used a single-pixel detector and a digital micromirror device to reconstruct intensity and lifetime images. A 1-D CNN architecture [27] was introduced to reduce the computational load for multi-exponential analysis using a similar branched structure. A multi-layer perceptron (MLP) method [28] was proposed for mono-exponential analysis with high spatial-resolution SPAD arrays. Another MLP [29] was reported combining maximum likelihood estimation algorithms and using fully connected layers to resolve bi-exponential decays. Moreover, another ANN technique [30] was introduced to fuse high-resolution fluorescence intensity and low-resolution lifetime images for wield-field FLIM systems. However, the training and inference of the ANNs are slow. Even with powerful GPUs, it usually takes a long training time (hours) to train a network. It is also time-consuming to retrain a model when the lifetime range is altered.

Pixel-wise lifetime recovery has been widely used since it is consistent with the sensor readout and computationally economical than 3-D algorithms. The extreme learning machine (ELM) [31] is an efficient algorithm to process 1-D signals for biological applications like electrocardiogram (ECG) and electroencephalogram (EEG) signals [32]. Inspired by related literature, we used ELM to reconstruct lifetimes from 1-D histograms using multi-variable regression. Contributions of the ELM-based lifetime inference approach are that:

1). It is data-driven without a back-propagation learning strategy. It achieves less training time than existing ANN methods, paving the way for fast online training on embedded hardware for FLIM.

2). It can resolve mono- and bi-exponential models widely employed in practical experiments, wherein the amplitude and intensity average lifetimes were investigated.

3). Reconstructed lifetime parameters from ELM are more accurate than fitting and non-fitting algorithms regarding synthetic and experimental data under different photon-counting conditions whilst maintaining fast computing speed.

This paper presents theory applying ELM to FLIM (Section 2), algorithms' comparisons regarding synthetic data with low photon count scenarios (Section 3), algorithms' comparisons regarding an incubated living cell under different levels of photon counts (Section 4).

## 2.  Apply ELM to FLIM

Due to ELM's superior capability of processing 1-D signals, we associated synthetic 1-D histograms with ELM regarding training and inferencing phases. We also illustrate the probabilistic model of photon arrivals of FLIM data and the artificial IRF based on TCSPC.

*2.1 ELM theory*

Conventionally, back-propagation is the gold standard to minimize object functions in most ANN architectures. ELM is theoretically a single-hidden layer feed-forward neural network (SLFN) that uses matrix inversion (or Moore-Penrose matrix inversion) and minimum norm least-square solution to train models. The training can be accelerated significantly compared with iterative back-propagation procedures whilst avoiding slow converges and over-fitting resulting from back-propagation. Assume H training samples (H pairs of vectors $\boldsymbol{x}_i = [x_{i1}, x_{i2}, ..., x_{im}]^T \in \mathbb{R}^m$ and $\boldsymbol{y}_i = [y_{i1}, y_{i2}, ..., y_{in}] \in \mathbb{R}^n$ are the $i$th input vectors and the $i$th target vectors, respectively. And suppose there are $L$ nodes in the single hidden layer; the output matrix of the hidden layer can be defined as:

$$A = \begin{bmatrix} \varphi(w_1 \cdot x_1 + b_1) & \cdots & \varphi(w_L \cdot x_1 + b_L) \\ \vdots & \ddots & \vdots \\ \varphi(w_1 \cdot x_H + b_1) & \cdots & \varphi(w_L \cdot x_H + b_L) \end{bmatrix}_{H \times L}, \quad (1)$$

where the $\varphi(\cdot)$ is the activation function, and usually, a *sigmoid* function can achieve a relatively good result. And $\boldsymbol{w}_l = [w_{l1}, w_{l2}, ..., w_{lm}]^T$ and $\boldsymbol{b}_l = [b_1, b_2, ..., b_L]^T$, $l=1, ..., L$. are randomly assigned weights and biases between the input nodes and the hidden layer before training. Say $\beta_l$ is the weighting connecting the $l$th hidden layer and output nodes, defined as:

$$\boldsymbol{\beta} = \begin{bmatrix} \beta_1^T \\ \vdots \\ \beta_L^T \end{bmatrix} = \begin{bmatrix} \beta_{11} & \cdots & \beta_{1n} \\ \vdots & \ddots & \vdots \\ \beta_{L1} & \cdots & \beta_{Ln} \end{bmatrix}_{L \times n}. \quad (2)$$

To learn the parameter matrix of $\boldsymbol{\beta}$ with a dimension of $L \times n$, the ridge loss function is widely adopted as:

$$\arg\min_{\beta \in \mathbb{R}^{L \times n}} \|A\boldsymbol{\beta} - Y\|^2 + \lambda \|\boldsymbol{\beta}\|^2, \quad (3)$$

where the $A$ is the matrix composed of the activation functions with dimensions $H \times L$; $Y$ is a matrix with dimensions $H \times n$ containing ground truth (GT) data:

$$Y = \begin{bmatrix} y_1^T \\ \vdots \\ y_H^T \end{bmatrix} = \begin{bmatrix} y_{11} & \cdots & y_{1n} \\ \vdots & \ddots & \vdots \\ y_{H1} & \cdots & y_{Hn} \end{bmatrix}_{H \times n}. \quad (4)$$

Through solving the loss function, we can obtain the matrix $\boldsymbol{\beta}$ by:

$$\hat{\boldsymbol{\beta}} = (A^T A + \lambda I)^{-1} A^T Y, \quad (5)$$

where $I$ is an identity matrix with dimensions $L \times L$, the hyperparameter $\lambda$ helps obtain a reliable result when the matrix $A^T A + \lambda I$ is not full-rank.

## 2.2 TCSPC model for FLIM

Fluorescence emission can be modeled with mono- or multi-exponential decay functions. And a bi-exponential model can approximately deduce a signal following a multi-exponential decay. Therefore, we focus on lifetime analysis from mono- and bi-exponential models in this work. Fluorescence functions can be adopted to formulate measured histograms containing multiple lifetime components and corresponding amplitude fractions. Therefore, for each pixel, the measured decay consisting of $K$ lifetime components is formulated as:

$$h(t) = IRF(t) * P \sum_{k=1}^{K} \alpha_k e^{-t/\tau_k} + n(t), \quad (6)$$

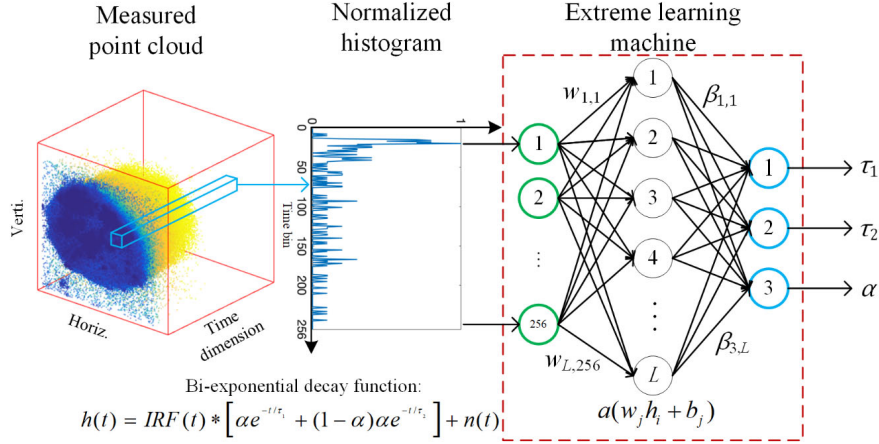

Fig. 1. ELM is used for lifetime analysis. The input data is a 1-D pixel-wise histogram from the raw point cloud that contains 256 time-bins. The histogram is fed into a single-hidden layer ELM, lifetime parameters ($\tau_1$, $\tau_2$, and $\alpha$) can be obtained from output nodes.

where the $IRF(\cdot)$ is the system's instrument response function, $P$ is proportional to the fluorescence intensity, $\tau_k$ is the $k$th lifetime component, $\alpha_k$ is the $k$th amplitude fraction, and $n(t)$ includes Poisson noise [33] and dark count rate of the sensor, $t = [1, 2, \ldots, T]$ is the time-bin index of the TCSPC module. As photon arrivals follow the Poisson distribution, with $C$ cycles of laser excitation, the ultimate distribution in one pixel can be derived as:

$$D \sim Poisson(C \int_0^T h(t)dt). \tag{7}$$

Based on this theoretical TCSPC model, we can generate training datasets for ELM. Synthetic curves correspond to column vectors in the input matrix $x$. Apart from multi-exponential decays, we define the amplitude-weighted lifetime $\tau_A$

$$\tau_A = \sum_{k=1}^{K} \alpha_k \tau_k, \tag{8}$$

and intensity-weighted average lifetime $\tau_I$

$$\tau_I = \frac{\sum_{k=1}^{K} \alpha_k \tau_k^2}{\sum_{k=1}^{K} \alpha_k \tau_k}. \tag{9}$$

to evaluate ELM.

### 2.3 Training data preparation

The training datasets contain 20,000 synthetic histograms, and GT lifetime parameters were generated to train the ELM network. Synthetic decays comply with Eq. (6). And the IRF curve is modelled via a Gaussian curve:

$$IRF(t) = e^{[-(t-t_0)^2 \cdot 4\ln 2 h^2 / FWHM^2]}, \tag{10}$$

where FWHM (0.1673 ns) is compatible with the two-photon FLIM system for FLIM measurements, $t_0$ (14$^{th}$) is the index of the peak, $h$ (0.039 ns) is the bin-width of the TCSPC system. And both mono- and bi-exponential decay models were generated for performance evaluation. Lifetime constants $\tau$ were set in [0.1, 5] ns for the mono-exponential decay model.

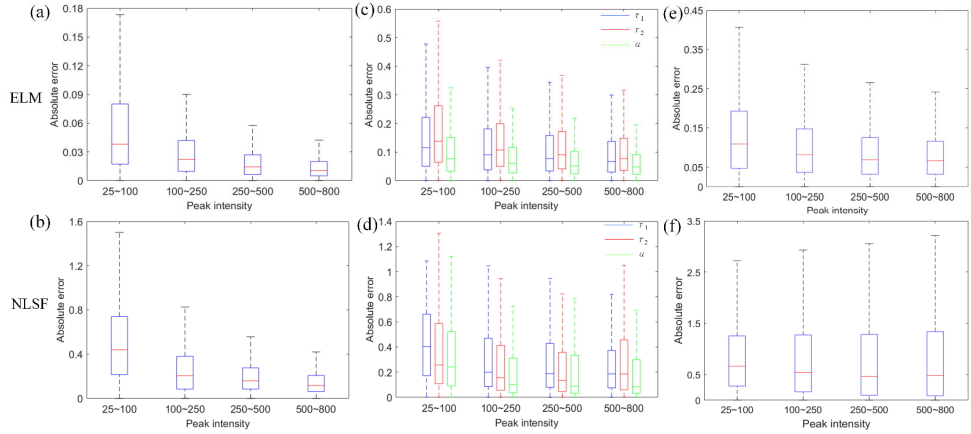

Fig. 2. Box plots of absolute error versus different peak intensity levels regarding testing datasets. (a) and (b) single lifetime estimations of mono-exponential decays from ELM and NLSF, respectively. (c) and (d) double lifetime estimations of bi-exponential decays from ELM and NLSF, respectively. (e) and (f) $\tau_A$ estimated by ELM and NLSF, respectively.

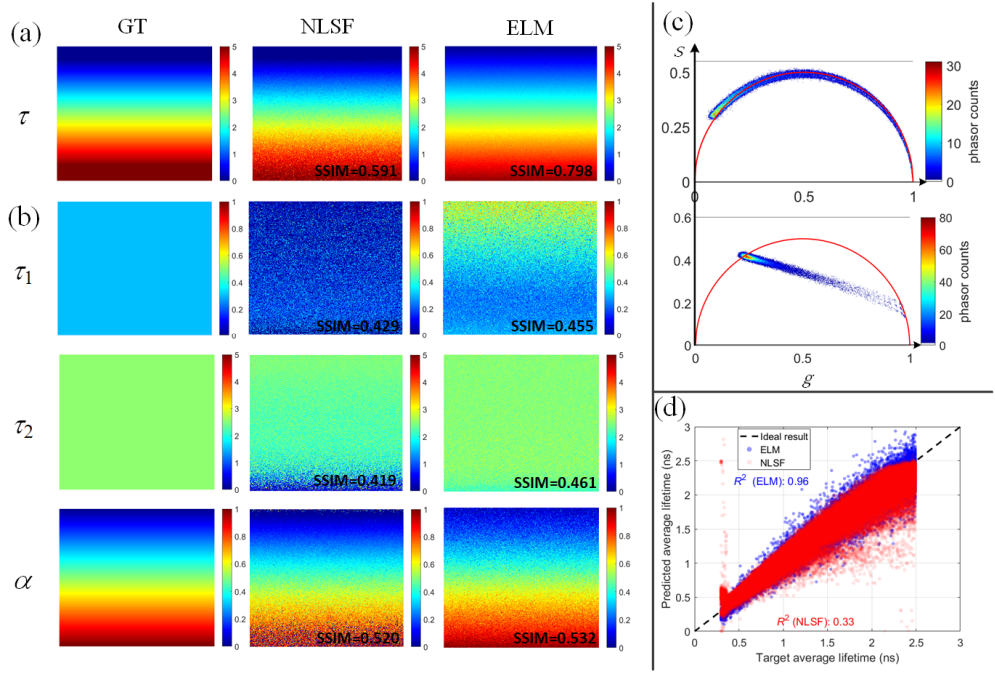

Fig. 3. Lifetime parameters' estimation results, photon counts for each pixel were randomly picked between 25 and 500. (a) The estimated single lifetime using a mono-exponential decay model, where $\tau \in [0.1, 5]$ ns from top to down in the image. (b) The two estimated lifetimes using a bi-exponential decay model where $\tau_1 = 0.3$ ns, $\tau_2 = 3$ ns, and $\alpha \in [0,1]$ from top to down. (c) Two phasor plots of ground truth distributions of (a) and (b). (d) Prediction accuracy and $R^2$ of $\tau_A$ from ELM and NLSF, with $\tau_1 = 0.3$ ns and $\tau_2 = 2.5$ ns, respectively.

And $\tau_1$, $\tau_2$ are set in $[0.1, 1]$, $[1, 3]$ *ns* for bi-exponential models. The structure of ELM is depicted in Fig. 1. Suppose the input vector is a pixel-wise histogram measured by a TCSPC system containing 256 time-bins in the inference phase. The number of output nodes depends on the number of lifetime components we defined in synthetic datasets. For instance, if the measured data is a bi-exponential decay model, the output layer should be configured as three nodes, namely, $\tau_1$, $\tau_2$, and $\alpha$. We can easily obtain average lifetimes from Eq. (8) and (9). All the histograms from the sensor are fed into the network sequentially; lifetime parameters can be obtained from output nodes pixel-by-pixel. The number of nodes in the hidden layer can be flexibly adjusted to achieve a trade-off between accuracy and computing time consumption.

## 3. Synthetic data analysis

$\tau_A$ and $\tau_I$ are used to estimate energy transfer for FRET or indicate fluorescence quenching behaviours [34]. This section compares NLSF, BCMM, and ELM to retrieve $\tau_A$ from bi-exponential decays. Likewise, we also compared NLSF, CMM, and ELM to reconstruct $\tau_I$. Besides, ELM was compared with existing ANNs for FLIM in terms of 1) the network scale and 2) training time. Multiple widely-used metrics (*F-Value*, SSIM, $R^2$, MSE) were adopted for performance evaluations.

### 3.1 Comparisons of individual lifetime components

As NLSF was usually adopted by previous studies [25], [27], [35], we compared the inference performances of ELM and deconvolution-based NLSF (implemented with *lsqcurvefit (·)* function in MATLAB using iterative Levenberg–Marquardt algorithm) in Fig. 2. 2,000 simulated testing datasets were generated for recovery for single and double lifetimes. Here we define the absolute error $\Delta g = |g - g_{est}|$, where $g = \tau_1$, $\tau_2$, $\alpha$, or $\tau_A$ and $g_{est}$ is the estimated $g$. $\Delta g_{ELM}$ and $\Delta g_{NLSF}$ are the absolute errors for ELM and NLSF. Fig. 2 (a) and 2 (b) show $\Delta g$ of ELM and NLSF for mono-exponential decays, respectively. $\Delta g$ decreases as the peak intensity increases, and $\Delta g_{ELM}$ is smaller than $\Delta g_{NLSF}$. Likewise, Fig. 2 (c) and (d) indicate $\Delta g$ plots for $g = \tau_1$, $\tau_2$, and $\alpha$, where $\Delta g_{ELM}$ is smaller than $\Delta g_{NLSF}$. Similarly, Fig. 2 (e) and (f) indicate ELM obtained a much more accurate $\tau_A$ than NLSF. Therefore, ELM can perform better than NLSF in mono- and bi-exponential decays.

Additionally, as shown in Fig. 3, we visually inspected estimated $\tau_1$, $\tau_2$, and $a$ based on pre-defined variables in synthetic 2-D images. We used the SSIM to evaluate reconstructed images. The 2-D lifetime images were reconstructed from a 3-D synthetic data cube (256×256×256, representing spatial and temporal dimensions). Fig. 3 (a) shows reconstructed 2-D images from mono-exponential decays with GT $\tau$ varying from 0.1 to 5 ns. Likely, Fig. 3 (b) shows estimated $\tau_1$, $\tau_2$, and $\alpha$ bi-exponential decays. Results obtained from ELM are more accurate than NLSF. Fig. 3 (c) shows the phasor plots of GT distributions of mono- (Fig. 3 (a)) and bi-exponential (Fig. 3 (b)) decays. From the phasor theory [36], cluster points of mono-exponential decays should locate on the semi-circle. For bi-exponential decays, two-lifetime components are indicated by the intersections of a fitted line and the semi-circle. We utilized $R^2$ defined as:

$$R^2 = 1 - \frac{\sum_{i=1}^{P}(\tau_A^i - \tau_{A\_GT}^i)}{\sum_{i=1}^{P}(\tau_A^i - \tau_{A\_Ave})}, \qquad (11)$$

to evaluate the estimation consistency, where $\tau_A^i$ is the predicted parameter, $\tau_{A\_GT}^i$ is the GT parameter, $\tau_{A\_Ave}$ is the average of GT parameters, $P$ is the number of simulated decay curves. As shown in Fig. 3 (d), the scatter plots show ELM is closer to GT, and NLSF shows more outliers. We further evaluated ELM and NLSF using the *F*-value defined as Eq. (12) [37] with synthetic mono- and bi-exponential decays.

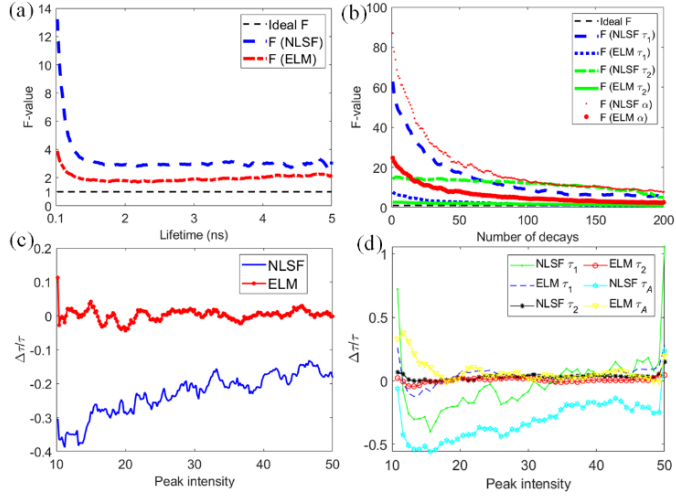

Fig. 4. (a) *F*-value for mono-exponential decays with a range [0.1, 5] ns. (b) *F*-values for bi-exponential decays with $\tau_1$, $\tau_2$, and $\alpha$) in the ranges [0.1, 1] ns, [1, 3] ns, and [0, 1], respectively. (c) and (d) bias per histogram for mono- and bi-exponential decays, respectively.

$$F = \frac{\delta x}{x} \cdot \sqrt{I}. \tag{12}$$

$F > 1$, and a lower $F$ means higher precision, where $I$ is the detected photon count, $\delta x$ is the standard deviation of the estimated lifetime parameter, and $x$ is the GT parameter. We generated 200 synthetic decays for given ranges of lifetimes and peak intensities in Fig. 4. Fig. 4 (a) shows the *F*-value of mono- exponential decays versus the lifetime in the range ~ [0.1, 5] ns. Fig. 4 (b) shows the *F*-value of bi-exponential decays versus $\tau_1$, $\tau_2$, and $\alpha$ in [0.1, 1] ns, [1, 3] ns, and [0, 1], respectively. We assigned 200 decays with a total photon count (< 2000) per synthetic histogram for both scenarios. Both figures show that ELM obtained a smaller *F* than NLSF, meaning ELM can achieve better precision. Furthermore, we defined the bias $\Delta\tau/\tau$ to evaluate ELM and NLSF versus the photon count. $\tau$ was set to 3.0 ns for mono-exponential decays. $\tau_1$, $\tau_2$, and $\alpha$ were set to 0.3 ns, 3.0 ns, and 0.5 for bi-exponential decays. Fig. 4 (c) shows that the bias of NLSF increases as the photon count increases, which is worse than ELM. And Fig. 4 (d) shows that the bias of ELM is smaller than NLSF, and ELM is more robust to varying photon counts. Moreover, NLSF is also sensitive to initial conditions of lifetime parameters [34]. The bias decreases when the initial conditions are closed to GT values, meaning that users need to have prior knowledge about the parameters to be extracted.

### 3.2 Comparisons of $\tau_A$

We evaluated ELM in estimating $\tau_A$ at various count conditions. As shown in Fig. 5 (a), we set three regions at three count levels changing $\tau_A$ from top to bottom. We called the three regions low, middle, and high counts hereafter. Fig. 5 (b) depicts the GT $\tau_A$. From Fig. 5 (c) and (d), ELM shows a more accurate $\tau_A$ image than NLSF, with ELM producing a smaller MSE than NLSF in each region. We also included the non- fitting BCMM [18] for the comparison due to its fast speed and capacity to resolve bi-exponential decays. From Fig. 5 (e), BCMM is not robust in low counts, outperforming NLSF in middle and high regions. Further, ELM obtained better results than BCMM. BCMM is less photon-efficient, and it is sensitive to the measurement window $T$ ($T$ should be larger than $5 \times \tau_2$, otherwise bias correction is needed [18]). Table 1 compares ELM with NLSF regarding the time consumption for inference (forward-propagation) tasks in Fig. 3 (a) and (b). NLSF resolving mono-exponential decays consumes more time than for bi-exponential decay models. In contrast, the analysis time of ELM is not affected by the number of lifetime components and substantially less than NLSF.

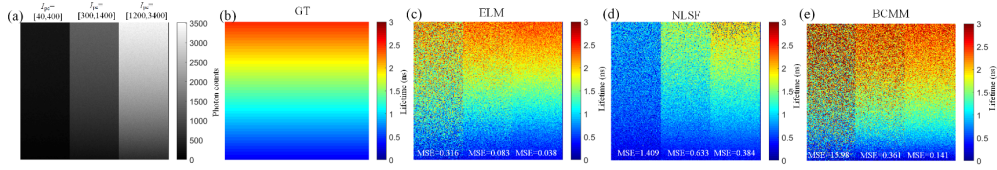

Fig. 5. (a) Intensity image of GT $\tau_A$ in exact ranges. $I_{pc}$ depicts total photon counts in one pixel. The range from 40 to 400 is viewed as low photon counts. (b) the GT $\tau_A$ lifetime image with the range ~ [0.3, 2.5] ns. (c), (d), and (e) reconstructed $\tau_A$ images from ELM, NLSF, and BCMM.

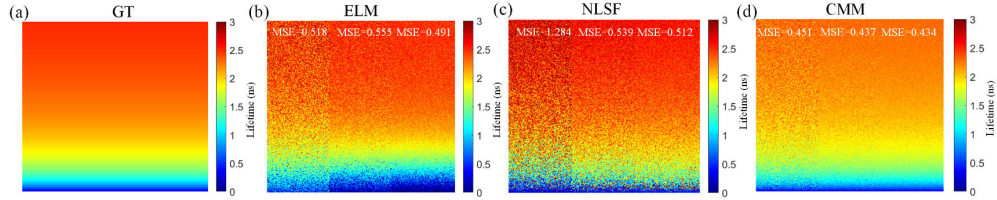

Fig. 6. (a) GT $\tau_I$ image in exact ranges. (b), (c), and (d) Reconstructed $\tau_I$ images from ELM, NLSF, and CMM for bi-exponential decays.

**Table 1. Time Consumption (Seconds) of NLSF And ELM for Inference Lifetime Parameters**

| Algorithm | Mono-exponential decay mode | bi-exponential decay mode |
|---|---|---|
| NLSF | 371.9 (sec) | 670.9 (sec) |
| ELM | 6.2 | 6.5 |
| CMM [17] | 1.9 | 1.9 ($\tau_I$) |
| BCMM [18] | - | 16.1 ($\tau_A$) |

### 3.3 Comparisons of $\tau_I$

CMM [17] achieves the fastest speed for intensity average lifetime analysis. We further compared CMM with ELM for $\tau_I$ reconstruction. As shown in Fig. 6, the result from ELM is better than NLSF but slightly worse than CMM. However, CMM is sensitive to and biased by the measurement window if bias correction is not included. Although CMM obtained a smaller overall MSE, the bias occurs as $\tau_I$ becomes longer. It agrees with the conclusion from the previous work [34], indicating that CMM causes misleading inference when there are multi-lifetime species in the field of view. Also, $\tau_I$ sometimes generates a shorter dynamic lifetime range than $\tau_A$ as $\tau_I$ cannot correctly distinguish clusters with different lifetimes, especially for strong FRET phenomena [5]. ELM and CMM can achieve a shorter processing time than NLSF and BCMM, as shown in Table 1. In this case, although ELM is slightly slower than CMM, the consumed time varies with the number of nodes in the hidden layer. Fig. 7 (a) shows training errors indicated by mean square errors (MAE) versus different numbers of nodes in the hidden layer. Here, the number of the hidden layer is set to 500 for both mono- and bi-exponential models as there was no apparent MAE decrease, and a moderate processing time was achieved, as shown in Fig. 7 (b). Moreover, we compared ELM with relevant ANNs for FLIM. Since

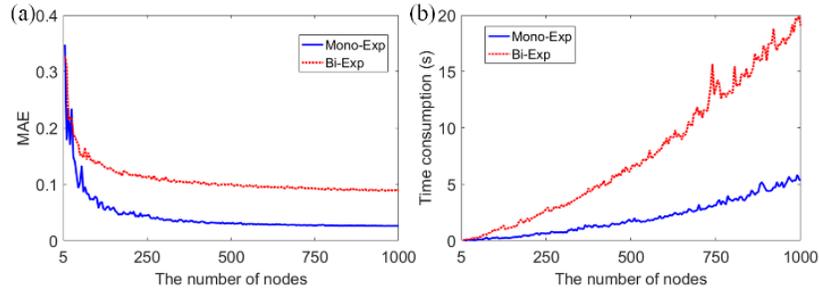

Fig. 7. (a), (b) Loss curves and time consumption versus different numbers of nodes in the hidden layer.

Table 2. Comparisons of Existing NN Architecture for Lifetime Estimation

| Algorithm | Training parameters | Hidden layer | Revolve multi-exp. decays | Training time |
|---|---|---|---|---|
| ELM | 205,600 | 1 | ✓ | 10.85 $s$ |
| FLI-NET [25] | 1,084,045 | 7 | ✓ | 4 $h$ |
| 1-D CNN [27] | 48,675 | 7 | ✓ | 23 $min$ |
| MLP [28] | 3,750,205 | 3 | ✗ | 38 $min$ |
| MLP [29] | 149,252 | 2 | ✓ | 4 $h$ |

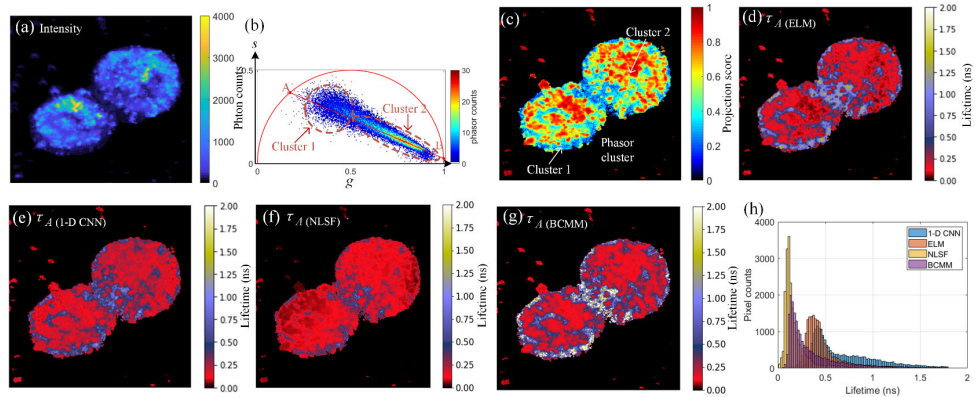

Fig. 8. Lifetime analysis of prostatic cells loaded with gold nanoprobes. (a) The intensity image, (b) phasor plot, and (c) phasor projection image. (d-g) $\tau_A$ restored by ELM, 1-D CNN, NLSF, and BCMM. (h) Lifetime histograms of ELM, 1-D CNN, NLSF, and BCMM.

ELM uses the Moore-Penrose matrix inversion strategy to learn parameters instead of back-propagation, it is much faster. As shown in Table 2, although ELM has more parameters than 1-D CNN [27], the training time is much shorter than the other existing studies [25], [28], and [29]. Many CNN hyperparameters should be fine-tuned, and batch normalizations should be implemented to avoid gradient vanishing [38]. In contrast, ELM's architecture is much simpler, and we just need to adjust the number of nodes in the hidden layer. Also, the efficient training process enables online training and is suitable for embedded hardware implementations [39]. ELM is highly reconfigurable to provide a flexible solution to balance the trade-off between computing complexity and accuracy. The evaluations of ELM and NLSF were conducted on MATLAB R2016a, 64-bit CPU (Intel Core i5-4200H @ 2.80GHz) with 8 GB memory. Notably, other studies in Table 2 used much more powerful GPU to train their models. Despite this, ELM still delivers the shortest training time.

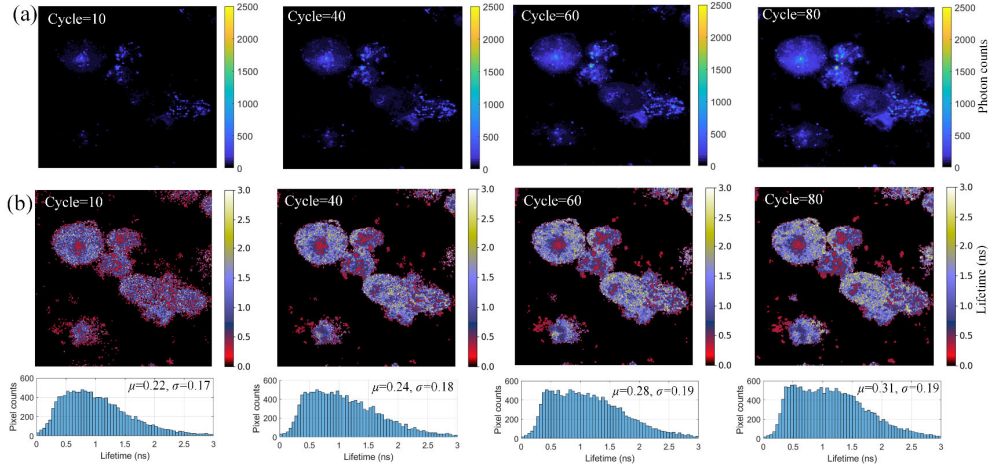

Fig. 9. (a) Intensity images with different scales of colorbars, scanning cycles were set to 10, 40, 60, and 80. Colorbars are unified. (b) $\tau_A$ images and pixel occurrence reconstructed by ELM at different cycles.

Based on the analysis of synthetic datasets, ELM is more robust to analyze mono- and bi-exponential decays than traditional NLSF methods. We will evaluate ELM using realistic experiment data in the next section.

## 4. Experiential FLIM data analysis

To investigate the feasibility of ELM for experimental FLIM data, we utilized living prostate cancer cells incubated with functionalized gold nanorods (GNRs). A commercial two-photon FLIM system was used to acquire raw 3-D data cubes. This section compares ELM with 1D-CNN, NLSF, and BCMM.

### 4.1 Experimental setup and sample preparation

We used the proposed ELM to analyze a living cellular sample acquired by a two-photon FLIM system. To achieve an efficient imaging contrast, prostate cancer cells were treated with GNRs functionalized with Cy5 labeled ssDNA [40]. GNRs have tunable longitudinal surface plasmon resonance and enable the interactions between the strong electromagnetic field and activated fluorophore in biological samples [41, 42]. Functionalizing GNRs with fluorophore labelled DNAs have been adopted to probe endocellular components [43, 44], like microRNA detections for human breast cancer or monitoring the intracellular level of metal ions in human serums. Here, prostate cancer cells were incubated with nanoprobe for 6 hours and washed three times with Phosphate-Buffered Saline (PBS). Cells were blended with 4% paraformaldehyde for 15min. After removing paraformaldehyde, cells were washed with distilled water three times. The two-photon FLIM platform consists of a confocal microscope (LSM 510, Carl Zeiss, Oberkochen, Germany) with 256×256 spatial resolution, where the scan module includes four individual PMTs. A TCSPC module (SPC-830, Becker & Hickl GmbH, Berlin, Germany) with 256 time-bin and 39 picosecond timing resolution was mounted on the microscope. A tunable femtosecond Ti: sapphire laser (Chameleon, Coherent, Santa Clara, USA) was configured with a repetition frequency 80MHz and 850 nm wavelength to excite the sample. The emission light was collected using a 60 × water-immersion objectives lens (numerical aperture = 1.0) and a 500-550 nm bandpass filter. One hundred scanning cycles were selected to prevent GNRs heating and obtain sufficient photons, where each cycle took three seconds.

*4.2 Algorithm evaluation*

Due to the strong two-photon photoluminescence property of GNRs, high optical discernibility can be observed between the GNRs and cell tissues [45]. Fig. 8 (a) shows the grey-scale intensity image of the sample, where the bright spots are GNRs. As the background pixels with fewer photon counts imply less useful information, they can be neglected during the analysis. In this case, a threshold (100 photon counts) was considered to neglect these pixels. As conventional data readout from TCSPC systems is pixel-by-pixel, accumulated histograms can be directly fed into the ELM without data conversion. The biological sample should be illuminated with a long acquisition time to achieve a high SNR to obtain a reliable reference. However, a long acquisition time can easily lead to photobleaching. The previous study [27] reported that a phasor projection image could alternatively serve as a reference image to identify autofluorescence and gold nanoprobes. After applying pixel filtering, two clusters representing autofluorescence of the cell and gold nanoprobes can be observed in the phasor plot shown in Fig. 8 (b). *Cluster 2* contains the majority of pixels with shorter lifetimes depicting gold nanoprobes. A fitted line was obtained by a linear regression fitting algorithm:

$$\arg\min_{a,b} \sum_{n=1}^{N} \|s_n - (ag_n + b)\|_2^2, \quad (13)$$

where $a$ and $b$ are slope and intercept of the fitted line, $g_n$ and $s_n$ are locations of pixels in the phasor domain. And the intersection points $A$ ($g_a$, $s_a$) and $B$ ($g_b$, $s_{2b}$) can be obtained accordingly. As shown in Fig. 8 (c), we employed the pixel-wise phasor score $\rho$ to generate a phasor projection image by computing:

$$\rho_n = [(g_n - g_2)(g_1 - g_2) + (s_n - s_2)(s_1 - s_2)] / D, \quad (14)$$

where $D$ is the Euclidean distance between $A$ and $B$, $n$ is the number of filtered pixels.

By comparing $\tau_A$ images obtained from ELM (Fig. 8 (d)), 1D-CNN (Fig. 8 (e)), NLSF (Fig. 8 (f)), and BCMM (Fig. 8 (g)), the image from NLSF shows obvious bias as we mentioned NLSF is sensitive to initial values and fails to converge sometimes. Given that the 1-D CNN [27] achieved high speed and accuracy, we compared ELM and 1-D CNN in terms of $\tau_A$ using the same training datasets. From Fig. 8 (d) and (e), ELM is in good agreement with 1D-CNN, and they showed similar distributions of pixel occurrences as shown in Fig. 8 (h). However, in Fig. 8 (g), the NLSF's result is significantly biased than the other three algorithms. This is because the deconvolution was involved in NLSF, causing non-convergent results due to dealing with ultra-short decays caused by gold nanoprobes. As mentioned, BCMM is not robust in varying ranges of photon counts; many pixels are out of the defined range (0 to 2 ns), as the white pixels shown in Fig. 8 (g). Nevertheless, BCMM is a fast algorithm that only took 6.53 seconds to reconstruct the image. The inference time of 1-D CNN on a GPU (NVIDIA GTX 850M) is 116.43 seconds, whereas ELM only consumed 1.73 seconds during inference on the CPU.

*4.3 Low counts scenarios*

Fragile tissues like retinas cannot be excited by laser for a long time. To avoid tissue damage and photobleaching caused by a long acquisition time, we investigated ELM's performance for data in low photon scenarios. We kept the experimental setup identical to Section 4.1. To acquire less emitted photons, we chose the field-of-view with fewer nanoprobes. Increased scanning cycles were set on the software. As the number of cycles increased, we changed the intensity threshold to guarantee sufficient pixels were saved. Fig. 9 (a) and (b) depict intensity and reconstructed $\tau_A$ images, respectively. The lifetime of cells and nanoprobes can be consistently reconstructed even the cycle decreases to 10. Notably, nanoprobes and boundaries of cells cannot be identified in intensity images with 10 and 40 cycles, yet lifetime images can restore the lifetime and reveal cell boundaries. Below each lifetime image in Fig. 9 (b), histograms of pixel occurrence were below $\tau_A$ images, showing means $\mu$ and standard deviations $\sigma$. There was no distinct shift of $\mu$ and $\sigma$ at different collection cycles, indicating that ELM is robust even at low counts.

## 5. Conclusion

In summary, we present an ELM architecture to accurately retrieve fluorescence lifetime parameters from mono- and bi-exponential decays. Both synthetic and realistic experimental FLIM datasets were employed to evaluate the proposed network. Our results show ELM outperforms fitting and non-fitting methods regarding synthetic datasets at different photon counts. And ELM can better identify NRs and cells and yield a comparable result to the 1-D CNN method. Since ELM does not need back-propagation to train the network, it is more flexible to reconfigure the network topology. Due to the potential online training property, it is promising to implement it on embedded hardware in the future, coupling with sensors and readout circuits to achieve fast on-chip training and inference. More FLIM applications relying on gold nanoparticles benefit from this study for cellular cancer diagnosis.

**Acknowledgments.** This work was supported in part by Datalab, Medical Research Scotland (MRS-1179-2017), Photon Force, Ltd., and BBSRC ( BB/V019643/1 and BB/K013416/1).

**Disclosures.** The authors declare that there are no conflicts of interest related to this paper.

**Data availability.** Data underlying the results presented in this paper are not publicly available at this time but may be obtained from the authors upon reasonable request.